\documentclass[prl,letterpaper,nofootinbib,floatfix,superscriptaddress,preprintnumbers,twocolumn]{revtex4}

\usepackage{color}

\usepackage{graphicx}

\usepackage{verbatim}
\usepackage{amsmath}
\usepackage{amssymb}
\usepackage{subfigure}
\usepackage{url}

\usepackage{multirow}
\usepackage{threeparttable}
\usepackage{paralist}

\usepackage{color}
\definecolor{darkgreen}{rgb}{0,0.5,0}

\newcommand{\be}{\begin{equation}}
\newcommand{\ee}{\end{equation}}

\newcommand{\IZ}{\mathbb{Z}}

\usepackage{soul}

\begin{document}

\title{Neutral Naturalness from the Orbifold Higgs}
 
\author{Nathaniel Craig} \email{ncraig@physics.ucsb.edu}
\affiliation{Department of Physics, University of California, Santa Barbara, CA 93106}
\affiliation{ Department of Physics and Astronomy, Rutgers University, Piscataway,
  NJ 08854}
  
\author{Simon Knapen} \email{smknapen@lbl.gov}
\affiliation{ Department of Physics and Astronomy, Rutgers University, Piscataway,
  NJ 08854}
  \affiliation{Berkeley Center for Theoretical Physics,
University of California, Berkeley, CA 94720}
\affiliation{Theoretical Physics Group, Lawrence Berkeley National Laboratory, Berkeley, CA 94720
}

\author{Pietro Longhi} \email{longhi@physics.rutgers.edu} 
\affiliation{ Department of Physics and Astronomy, Rutgers University, Piscataway,
  NJ 08854}

\date{\today}

\begin{abstract}
We present a general class of natural theories in which the Higgs is a pseudo-goldstone boson in an orbifolded gauge theory. The symmetry protecting the Higgs at low energies is an accidental global symmetry of the quadratic action, rather than a full continuous symmetry. The lightest degrees of freedom protecting the weak scale carry no Standard Model (SM) quantum numbers and interact with visible matter principally through the Higgs portal. We find that the twin Higgs is the simplest example of an orbifold Higgs. This opens the door to the systematic study of ``neutral naturalness'': natural theories with SM-neutral states that are as yet untested by the LHC.

\end{abstract}

\preprint{RU-NHETC-{2014-15}}

\maketitle

\section{Introduction}

The discovery of an apparently elementary Standard Model (SM)-like Higgs at the Large Hadron Collider (LHC) \cite{Aad:2012tfa, Chatrchyan:2012ufa} has heightened the urgency of the hierarchy problem, yet comprehensive null results in searches for new physics have called the naturalness of the weak scale increasingly into question. 
The tension between naturalness and null results is largely driven by a folk theorem regarding top partners: In symmetry solutions to the hierarchy problem, the symmetry protecting the Higgs from sensitivity to high mass scales typically commutes with the SM gauge group, giving rise to top partners that are charged under QCD. The precise character of the top partners may vary, ranging from scalar top partners in the case of supersymmetry to vector-like fermionic top partners in the case of global symmetries, but in either case their QCD quantum numbers guarantee a large production cross section at the LHC. Limits on these top partners are now approaching 700-800 GeV for both scalar and fermionic top partners \cite{Aad:2014kra, Chatrchyan:2013xna, Aad:2014efa, Chatrchyan:2013uxa}, imperiling the apparent naturalness of the weak scale. While there is still room left for light colored top partners, the generic reach of LHC limits has created substantial tension.

Before abandoning the naturalness principle, it is reasonable to wonder whether all natural theories have been systematically explored -- and, in particular, whether there are general classes of models preserving naturalness through symmetries for which the top partner folk theorem does not apply. There is an encouraging proof of principle -- the twin Higgs model \cite{Chacko:2005pe}, for which the lowest-lying degrees of freedom required for naturalness are neutral under the SM -- but it remains unclear whether this is an isolated point in theory space or an example of a generic phenomenon based on symmetries.  

In this paper we identify an expansive class of new theories that preserve the naturalness of the weak scale without predicting new light degrees of freedom charged under the Standard Model. We exploit the fact that the correlation functions of orbifold daughter theories are identical to those of the parent theory in the large-$N$ limit \cite{Kachru:1998ys,Bershadsky:1998mb,Kakushadze:1998tr, Bershadsky:1998cb, Schmaltz:1998bg}.\footnote{Orbifolds have been fruitfully used to construct natural theories with scalar \cite{Burdman:2006tz, Craig:2014fka} or fermionic \cite{Poland:2008ev, Cai:2008au} top partners charged under the SM electroweak group but neutral under QCD. In contrast, here we construct a general class of models where all additional states in the low energy theory are entirely neutral under the Standard Model.} 
The key observation is that orbifolds of continuous symmetries may produce daughter theories featuring an accidental symmetry of the quadratic action.
As such, the Higgs can be identified as a pseudo-goldstone boson of the spontaneously broken accidental symmetry. Radiative stability is guaranteed by appropriate pseudo-goldstone couplings to fermions charged under gauge sectors other than the Standard Model. 

We outline the essential framework of the orbifold Higgs and provide several illuminating examples. Surprisingly, we find that the twin Higgs model {\it is} the simplest example of an orbifold Higgs, and the orbifold structure can explain all of the features required by a successful twin Higgs model. In the interest of brevity, we leave many technical details of the orbifold procedure and detailed model-building to a companion paper \cite{CKL}.

\section{Orbifold field theory}

We begin by briefly reviewing the structure of field theory orbifolds, as discussed in further detail in \cite{Schmaltz:1998bg, CKL}. To orbifold a parent symmetry $G$ by a discrete group $\mathcal{G}$, we first embed $\mathcal{G}$ into $G$ using the regular representation embedding. We obtain the daughter theory by studying how $\mathcal{G}$ transformations act on fields charged under $G$ and projecting out all states not invariant under those transformations.\footnote{For our purposes it suffices to consider such orbifolds at the level of field theory, but they can be realized naturally in the compactification of higher-dimensional theories with non-trivial boundary conditions. In this context the field theory orbifold provides the untwisted spectrum, while dimensional reduction gives rise to additional towers of states in the twisted spectrum around the scale of the inverse compactification radius.} Such orbifold daughter theories enjoy the surprising property that their correlation functions are identical to those of the parent theory in the large-$N$ limit \cite{Kachru:1998ys,Bershadsky:1998mb,Kakushadze:1998tr, Bershadsky:1998cb, Schmaltz:1998bg}. 

As an example, consider a toy model consisting of an $SU(N \Gamma)$ parent gauge theory with a field $H$ transforming as a bi-fundamental under the gauge symmetry and an $SU(\Gamma)$ flavor symmetry. Taking the orbifold by $\IZ_\Gamma$ gives rise to a daughter theory with gauge group $SU(N)^\Gamma \times U(1)^{\Gamma -1}$, while the parent bi-fundamental $H$ decomposes into $\Gamma$ $SU(N)$ fundamentals, one charged under each $SU(N)$ factor of the daughter symmetry. More generally, when orbifolding by a $\Gamma$-dimensional discrete group $\mathcal{G}$ with number of irreducible representations $n_{\mathcal{G}}$, the parent symmetry is broken into a product of daughter symmetries of relative dimension $d_\alpha$: 
 \begin{equation}
 SU(\Gamma N) \to \left( \prod_{\alpha = 1}^{n_{\mathcal{G}}} SU(d_\alpha N) \right) \times \left( U(1) \right)^{n_{\mathcal{G}}-1}
 \end{equation} 
 where the $d_\alpha$ factors and {decomposition of matter multiplets both depend in detail on the group $\mathcal{G}$}. For our purposes, it is critical to understand the scaling of couplings in the daughter theory relative to the parent symmetry, which are dictated by the orbifold projection and canonical normalization of daughter states. Specifically, gauge couplings $g_\alpha$ and yukawa couplings $y_\alpha$ of a daughter theory sector with gauge group $SU(d_\alpha N)$ are rescaled relative to the parent couplings $g, y$ by $g_\alpha = g / \sqrt{d_\alpha}$ and $y_\alpha = y / \sqrt{d_\alpha}$, respectively, while mass terms $m_\alpha$ and quartic couplings $\lambda_\alpha$ are not rescaled, $m_\alpha = m$ and $\lambda_\alpha = \lambda$ \cite{Schmaltz:1998bg, CKL}. This will play a crucial role in the radiative stability of orbifold Higgs models. 

\section{The orbifold Higgs}

\begin{figure}[t] 
   \centering
   \includegraphics[width=3in]{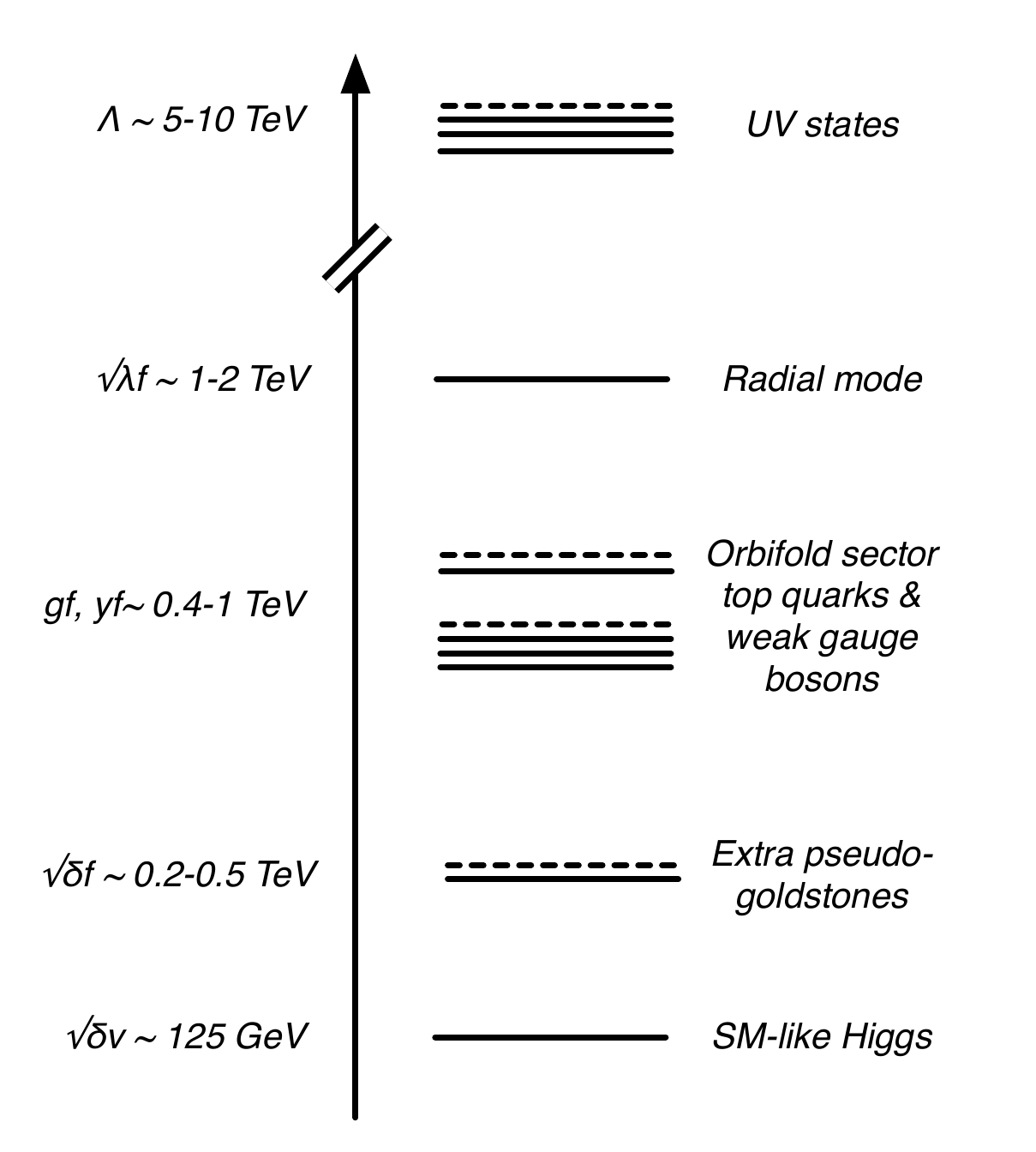} 
   \caption{Schematic scales in an orbifold Higgs model.}
   \label{fig:scales}
\end{figure}

In a viable orbifold Higgs theory, we envision a cutoff scale $\Lambda \sim 10$ TeV above which the parent symmetry is manifest, and below which the parent theory is reduced to the daughter theory. If the orbifold is realized geometrically, this scale corresponds roughly to the inverse compactification scale of extra dimensions. As with any global symmetry protection for the Higgs, a mechanism such as supersymmetry, compositeness, or large extra dimensions protects the low-energy degrees of freedom from sensitivity to scales above the cutoff. At the level of the field theory orbifold, the parent theory consists minimally of an appropriate parent symmetry and corresponding parent fields $H, Q, U$ -- which will give rise to Higgs and top multiplets in the daughter theory -- and is orbifolded by a discrete group $\mathcal{G}$ to obtain a daughter theory that includes a copy of $SU(3) \times SU(2)$ as well as other orbifold sectors. The daughter theory contains various other Higgs and top multiplets distributed among the other orbifold sectors. 
The quadratic potential for the Higgs multiplets in the daughter theory will respect a continuous symmetry originating from the {\it parent} theory.
When the Higgs multiplets acquire vevs, this accidental continuous symmetry will be spontaneously broken with a collective order parameter $f \sim$ TeV, leading to a radial mode of the accidental symmetry breaking and some number of uneaten pseudo-goldstones. With an appropriate perturbation of the vacuum to ensure $v \ll f$, the pseudo-goldstone aligned with $v$ may be identified with the SM-like Higgs. The quartic coupling $\delta$ of the SM-like Higgs is radiatively generated from SM gauge and yukawa couplings. There may be additional pseudo-goldstones aligned with other orbifold sectors that are parametrically heavier, with masses of order $\sim \sqrt{\delta} f$. The role of the top partner is played by an admixture of top quarks in the other orbifold sectors with masses of order $y_t f$, with couplings dictated by integrating out the radial mode. Similarly, the roles of weak gauge partners are played by an admixture of gauge bosons in the other orbifold sectors with masses of order $g f$. The relevant scales and states are sketched in Fig.~\ref{fig:scales}.
 
For the sake of concreteness, consider the simplest example of an orbifold Higgs where $\mathcal{G} = \IZ _{2}$ is embedded into the gauge symmetries through the regular representation. For the time being, we will focus on the degrees of freedom most relevant for naturalness of the weak scale, namely the Higgs, third generation fermions, top yukawa, QCD, and weak gauge interactions.\footnote{We will reserve discussion of $U(1)$ factors, leptons, light generations, and the bottom yukawa for the next section. Consequently, the toy models illustrated here are anomalous, but anomaly cancellation succeeds in UV completions including down-type quarks and leptons.} The parent symmetry consists of a gauge group $SU(6) \times SU(4)$. The matter consists of a scalar $H$ and two fermion multiplets $Q$ and $U$, transforming under the gauge group and an $SU(2)$ global symmetry as
\begin{equation}\label{eq:mother-table}
\begin{array}{c|cc|c}
&SU(6) &SU(4)&SU(2)\\\hline
H&1& \square& \overline{\square}\\
Q& \square& \overline{\square}&1\\
U& \overline{\square}&1&\square
\end{array}
\end{equation}
These symmetries allow a parent top yukawa, Higgs mass, and Higgs quartic coupling of the form
\begin{equation} \label{eq:ppot}
V_{p} = y H Q U - m^2 |H|^2 +  \lambda (|H|^2)^2 
\end{equation} 
where we neglect a single-trace quartic operator with some mischief aforethought.
The daughter theory is obtained by taking the orbifold $[SU(6)\times SU(4)] / \IZ_2$. The resulting gauge group is $[SU(3)\times SU(2)]_A \times [SU(3) \times SU(2)]_B \times U(1)^2$. We will set aside the abelian factors for the time being (or lift them without consequence via the Stueckelberg mechanism); on their own they are unsuitable for identification with hypercharge, but may be mixed with other $U(1)$ generators to suit this purpose. The matter consists of $H_A, Q_A, U_A$ transforming as a Higgs doublet, left-handed quark doublet, and right-handed quark of $SU(3)_A \times SU(2)_A$, and likewise $H_B, Q_B, U_B$ transforming analogously under $SU(3)_B \times SU(2)_B$. There are no matter fields transforming under both $A$ and $B$ groups. The relative dimensions of the daughter groups are all identical, $d_\alpha = 1$, so that the gauge, yukawa, mass, and quartic couplings inherited in the daughter theory are the same as those of the parent. In particular, the $SU(3)_A$ and $SU(3)_B$ gauge couplings are identical, as are those of $SU(2)_{A}$ and $SU(2)_{B}$, while the interactions of the daughter theory take the form
\begin{eqnarray} \nonumber
V_d &=& y H_A Q_A U_A + y H_B Q_B U_B - m^2 (|H_A|^2 + |H_B|^2) \\ \label{eq:tree} &+& \lambda( |H_A|^2 + |H_B|^2)^2
\end{eqnarray} 
Note that these are precisely the interactions present in the twin Higgs \cite{Chacko:2005pe}. Whereas in the twin Higgs model the fundamental symmetry is posited to be of the form ${\rm SM}_A \times {\rm SM}_B \times \IZ_2$, here the daughter symmetry  ${\rm SM}_A \times {\rm SM}_B \times \IZ_2$ is merely a residual resulting from the orbifold $[SU(6)\times SU(4)] / \IZ_2$. The orbifold also provides an $SU(4)$-symmetric quartic in the daughter theory if the parent theory quartic is dominated by double-trace terms as in (\ref{eq:ppot}). This provides a guide for specific UV completions, as an approximately $SU(4)$-symmetric quartic is required in the twin Higgs but cannot be explained by the twin $\IZ_2$ symmetry alone.

The power of the orbifold Higgs is manifest in the structure of radiative corrections to the Higgs multiplets in the daughter theory. As a proxy for the impact of radiative corrections from mass thresholds at higher scales, the quadratic part of the one loop effective potential with uniform cutoff $\Lambda$ takes the form\footnote{In a UV complete theory the cutoff $\Lambda$ will be replaced by physical mass thresholds respecting the parent gauge symmetry, justifying the choice of equal cutoffs in the $A$ and $B$ sectors and rendering (\ref{eq:cw}) a useful proxy. Above the scale of these thresholds, the usual full UV completions such as supersymmetry, compositeness, or large extra dimensions will protect against corrections from yet higher scales.}
\begin{equation} \label{eq:cw}
V^{(1)}_d \supset \frac{\Lambda^2}{16 \pi^2} \left( - 6 y^2 + \frac{9}{4} g^2 + 10 \lambda \right)   \left(|H_A|^2 + |H_B|^2 \right)
\end{equation}
Note that these radiative corrections are proportional to an $SU(4)$ invariant $|H_A|^2 + |H_B|^2$, despite the fact that the daughter theory does not possess a continuous $SU(4)$ symmetry; it is merely an accidental symmetry of the quadratic action. 
This is the essential feature of the orbifold Higgs: the heritage of a parent theory is an accidental symmetry of the quadratic action in the daughter theory. 
As such, when the Higgs multiplets of the daughter theory acquire vacuum expectation values, the physical SM-like Higgs $h$ may be identified with an uneaten pseudo-goldstone boson of the spontaneously broken accidental symmetry. This ensures that the potential for $h$ is independent of $SU(4)$ invariants, and thus insensitive to radiative corrections from higher scales.

At one loop there are also radiatively generated quartics for each doublet that break the accidental $SU(4)$ symmetry, namely
\begin{equation}\label{eq:cwquartics}
V^{(1)}_d \supset \frac{3}{16 \pi^2} y^4 |H_A|^4 \log \left( \frac{\Lambda^2}{y^2 |H_A|^2} \right)  + A\rightarrow B + \dots
\end{equation}
where $\dots$ includes numerically subleading corrections proportional to $g^4$ as well as $SU(4)$-preserving radiative quartics proportional to $\lambda$. The vacuum structure of the tree-level potential (\ref{eq:tree}) perturbed by radiative corrections (\ref{eq:cw}) and (\ref{eq:cwquartics}) gives $|\langle H_A \rangle|^2 = |\langle H_B \rangle|^2 = f^2 / 2$, so that the goldstone mode of spontaneous $SU(4)$ breaking is equally aligned with the two sectors. In order to identify the goldstone mode with an SM-like Higgs, the vev must be rendered asymmetric by either a soft (mass) or hard (quartic) breaking of the residual $\IZ_2$ symmetry in the daughter theory. These terms explicitly violate the parent symmetry, but may be induced in a geometric UV completion. Either perturbation can result in a vacuum with $\langle H_A \rangle \ll \langle H_B \rangle$, such that the uneaten goldstone of spontaneous $SU(4)$ breaking is primarily aligned with the physical fluctuation around $\langle H_A \rangle \equiv v$ -- i.e., can be identified with the SM-like Higgs -- and the radial mode is primarily aligned with the physical fluctuation around $\langle H_B \rangle \sim f$. The role of the top partner can be understood by integrating out the radial mode, in which case the SM-like Higgs inherits precisely the couplings to the $B$-sector top quarks required of fermionic top partners, yet these top partners are entirely neutral under the SM gauge group. In this manner, the Higgs is a pseudo-goldstone boson of an orbifolded $SU(4)$ gauge symmetry and the degrees of freedom guaranteeing radiative stability of the weak scale are neutral under the SM. 

The orbifold Higgs procedure enables the generalization of the twin Higgs mechanism to arbitrary discrete symmetries $\mathcal{G}$, illuminating an extensive class of theories where the naturalness of the weak scale is preserved by degrees of freedom neutral under the Standard Model. Perhaps the simplest generalization is $\mathcal{G} = \IZ_\Gamma$. Here the parent gauge group is $SU(3 \Gamma) \times SU(2 \Gamma)$, and the orbifold daughter symmetry is $[SU(3) \times SU(2)]^\Gamma \times U(1)^{\Gamma - 1}$. As in the case of $\Gamma = 2$, the Standard Model may be identified with one $SU(3)\times SU(2)$ factor, with the remaining $\Gamma - 1$ factors comprising a set of identical orbifold sectors. The theory possesses an accidental $SU(2 \Gamma)$ symmetry spontaneously broken by the vevs of the Higgs multiplets charged under the $\Gamma$ $SU(2)$ factors in the daughter theory. In this case there are $\Gamma - 1$ uneaten goldstones, of which one may be identified with the SM-like Higgs and the others are primarily aligned with other sectors. The Higgs is a pseudo-goldstone of an accidental $SU(2 \Gamma)$ symmetry at the quadratic level of the daughter action. The protection by the top partners is now provided by a linear combination of operators involving top quarks in the $\Gamma - 1$ orbifold sectors, with couplings again dictated by integrating out the radial mode. 

More interesting examples arise when $\mathcal{G}$ is nonabelian. Consider, for example, $\mathcal{G} = S_3$. In this case the parent gauge group is $SU(18) \times SU(12)$, and the orbifold daughter is $[SU(3) \times SU(2)]_A \times [SU(3) \times SU(2)]_B \times [SU(6) \times SU(4)] \times U(1)^4$. The two $SU(3) \times SU(2)$ sectors each inherit a Higgs doublet $H_{A,B}$, left-handed quark doublet $Q_{A,B}$, and right-handed quark $U_{A,B}$ as before, while the $SU(6) \times SU(4)$ sector inherits a pair of Higgs fourplets $H_{C,C'}$, a single left-handed quark fourplet $Q_C$, and two right-handed quarks $U_{C,C'}$. Now the SM gauge sector may be identified with one of the $SU(3) \times SU(2)$ copies and the SM-like Higgs with a pseudo-goldstone of the spontaneously broken accidental $SU(12)$ symmetry. The two $SU(3)\times SU(2)$ sectors have relative dimension $d_\alpha = 1$, and so inherit couplings directly from the parent theory, while the $SU(6) \times SU(4)$ sector has relative dimension $d_\alpha = 2$ and inherits appropriately rescaled yukawa and gauge couplings. The rescaling of couplings guarantees that the one-loop radiative corrections to the Higgs multiplets retain the $SU(12)$ form up to $1/N$ corrections, 
\begin{eqnarray} \label{eq:cws3}
V^{(1)}_d \supset \frac{\Lambda^2}{16 \pi^2} \left( - 6 y^2 + \frac{9}{4} g^2 + 26 \lambda \right)  \left( \sum_{I = A,B,C,C'} |H_I|^2 \right) \\ \nonumber
+ \frac{\Lambda^2}{16 \pi^2}\left(\frac{9}{16} g^{2} \right) \left( |H_C|^2 + |H_{C'}|^2 \right)
\end{eqnarray}
precisely as required by the orbifold correspondence. The residual cutoff sensitivity of the Higgs multiplets charged under the $SU(4)$ sector is modest for a cutoff in the tens of TeV. The protection from the top partners in this theory originates from a linear combination of operators involving top quarks in the other $SU(3)\times SU(2)$ sector and the $SU(6)\times SU(4)$ sector.

\section{Discussion \& Conclusions}

Thus far we have focused on toy models with a field theoretic orbifold, which identifies the relevant degrees of freedom at low energies. Such orbifolds are typically realized geometrically by the reduction of higher-dimensional theories subject to non-trivial boundary conditions, which guarantees the form of the field theory orbifold and provides a natural mechanism for including couplings and states that violate the parent symmetry by localization at lower-dimensional defects.  The $\mathcal{G} = \IZ_2$ orbifold can be naturally realized in five dimensions, whereas more complicated orbifolds such as $\IZ_\Gamma$ or $S_3$ require additional extra dimensions. Alternately, the same physics may be reproduced entirely in four dimensions using dimensional deconstruction \cite{ArkaniHamed:2001ca}.

Schematically, the parent symmetry is a gauge symmetry of the higher-dimensional bulk, with boundaries respecting the gauge groups of the daughter symmetry (but not necessarily the residual discrete symmetries). The Higgs and third-generation top multiplets should be identified with bulk states, while degrees of freedom that are irrelevant for naturalness (such as first- and second-generation Standard Model fermions) or do not fit into the parent symmetry can be localized on the lower-symmetry boundaries. Down-type quarks and leptons of the third generation could be either bulk- or boundary-localized; the former requires two Higgs doublets in the bulk to allow Yukawa couplings, while the latter is compatible with a single bulk Higgs multiplet. Localizing states on defects in geometric orbifolds liberates the orbifold Higgs from a major drawback of the twin Higgs scenario, namely the presence of light generations in the twin sector. These are irrelevant for naturalness but remain in thermal equilibrium with the Standard Model down to low temperatures and lead to potentially prohibitive contributions to $N_{\rm eff}$. They are present in twin Higgs models with an exact $\IZ_2$ symmetry but can be eliminated entirely in geometric orbifolds by localizing light generations on the boundaries. 

We have also set aside discussion of hypercharge embeddings, for which there are many possible options. Perhaps the simplest is to begin with a universal $U(1)$ in the parent theory such that all daughter sectors are charged under the same $U(1)$. This implies that all hidden sector fields will have hypercharge assignments identical to those of their Standard Model partners, giving constraints from null searches for $Z'$ gauge bosons, heavy stable charged particles, and exotic particles coupling to the $Z$ boson. A more straightforward option is to obtain independent hypercharge factors for each daughter sector by extending the parent groups from $SU(N)$ to $U(N)$. For example, the orbifold $U(6)\times U(4)/\IZ_2$ gives an $[SU(3)\times SU(2)]^2 \times U(1)^4$ daughter theory, in which two linear combinations of $U(1)$ factors can be identified with hypercharge factors $U(1)_{A,B}$, and the remaining two linear combinations lifted by the Stueckelberg mechanism. Or, as is often the case in string theory realizations of the Standard Model gauge group, hypercharge may be obtained from embedding in a low-scale ``unified'' group such as Pati-Salam $SU(4) \times SU(2) \times SU(2)$ \cite{Pati:1974yy} or $SU(3) \times SU(3) \times SU(3)$ trinification \cite{Glashow:1984gc}. In each case the parent symmetries can be extended appropriately to give rise to the unified group in the daughter theory, and these  unified groups may be broken by orbifolding in the usual manner without giving rise to dimension-6 proton decay. An even more radical possibility in the context of geometric UV completions would be to charge both bulk and defect-localized states under a defect-localized $U(1)$ hypercharge factor along the lines of \cite{Kehagias:2005be} -- so that only the SM sector would have a hypercharge gauge boson at low energies -- but it is unclear if this proposal can be consistently realized for bulk matter fields charged under both bulk and defect-localized gauge symmetries.

There are numerous future directions. Here we have sketched the essential structure of orbifold Higgs models and illustrated features of UV complete models, but it would be worthwhile to construct explicit UV completions and explore their phenomenology. We have also restricted our attention here to orbifolds of gauge symmetries in the parent theory; it would be extremely interesting to extend our analysis to simultaneous orbifolds of gauge and $R$-symmetries in supersymmetric models, leading to the generalization of models such as folded supersymmetry \cite{Burdman:2006tz}. More generally, the orbifold Higgs demonstrates that entirely novel theories for the weak scale may be generated by the appropriate reduction of more familiar continuous symmetries. We have focused on orbifolding by the regular embedding of a discrete symmetry, but in principle there are many other possibilities, including orbifolds using alternate embeddings of the discrete symmetry; orientifolds; or more exotic defects arising in the context of string theory.

\acknowledgments We thank Aria Basirnia, Zackaria Chacko, Tony Gherghetta, Roni Harnik, Kiel Howe, Andrey Katz, Yasunori Nomura, Duccio Pappadopulo, Michele Papucci, Stuart Raby, Michael Ratz, Martin Schmaltz, Raman Sundrum, and especially Matt Strassler for useful conversations. This work was supported in part by DOE grants SC0010008, ARRA-SC0003883, and DE-SC0007897. N.C. acknowledges support from the Aspen Center for Physics and NSF grant 1066293 where this work was partially completed. This manuscript has been authored by an author (SK) at Lawrence Berkeley National Laboratory under Contract No. DE-AC02-05CH11231 with the U.S. Department of Energy. The U.S. Government retains, and the publisher, by accepting the article for publication, acknowledges, that the U.S. Government retains a non-exclusive, paid-up, irrevocable, world-wide license to publish or reproduce the published form of this manuscript, or allow others to do so, for U.S. Government purposes.

\bibliography{orbibib}

\end{document}